\documentclass{appolb}
\usepackage{graphicx}
\usepackage{xcolor}
\usepackage{wrapfig}
\usepackage{amsmath}
\usepackage{amssymb}
\usepackage{multirow}
\usepackage{slashed}
\usepackage{cite}

\newcommand{\pbp}{\langle\bar{\psi}\psi\rangle}

\begin{document}
\title{QCD phase structure in strong magnetic fields%
\thanks{Presented in the online workshop on Criticality in QCD and Hadron Resonance Gas, July 29-31, 2020, Wroclaw, Poland.}%
}
\headtitle{QCD phase structure in strong magnetic fields}
\headauthor{H.-T. Ding, S.-T. Li, Q. Shi, A. Tomiya, X.-D. Wang, Y. Zhang}
\author{
	H.-T. Ding$^{\rm a}\thanks{Speaker}$, S.-T. Li$^{\rm b,a}$, Q. Shi$^{\rm a}$, A. Tomiya$^{\rm c}$,\\ X.-D. Wang$^{\rm a}$, Y. Zhang$^{\rm a}$
\address{
	$^{\rm a}$Key Laboratory of Quark $\&$ Lepton Physics (MOE) \\and Institute of Particle Physics, \\
	Central China Normal University, Wuhan 430079, China\\
	$^{\rm b}$Institute of Modern Physics, Chinese Academy of Sciences, \\Lanzhou 730000, China \\
	$^{\rm c}$RIKEN BNL Research center, Brookhaven National Laboratory, \\Upton, NY, 11973, USA 
}}

\maketitle
\begin{abstract}
In this proceedings we discuss the natural connection between the reduction of neutral pion mass in the vacuum, and the magnetic catalysis as well as the reduction of transition temperature in the external magnetic field. We also present the first results on fluctuations of and correlations among conserved charges in strong magnetic fields from lattice QCD computations.

\end{abstract}
  
\section{Introduction}
There have been a lot of interesting results obtained from studies on the QCD thermodynamics in the external magnetic field~\cite{Kharzeev:2007jp,DElia:2012ems,Kharzeev:2015znc,Endrodi:2014vza}. Among these results two well-known phenomena are the reduction of transition temperature $T_{pc}$ and the (inverse) magnetic catalysis~\cite{Andersen:2014xxa}. It was expected that the transition temperature should increase due to the magnetic catalysis observed at zero temperature~\cite{DElia:2010abb,Shovkovy:2012zn}. However, it turns out that the transition temperature actually decreases as the magnetic field strength $eB$ grows~\cite{Bali:2011qj}. The incorrect increasing behavior of $T_{pc}$ in $eB$ is mainly due to the large lattice cutoff effects in the standard staggered discretization schemes~\cite{DElia:2010abb,Bali:2011qj,Ding:2020inp}. And in QCD with physical values of pions the reduction of $T_{pc}$ accompanies with a non-monotonous behavior of chiral condensate near $T_{pc}$~\cite{Ilgenfritz:2013ara,Bornyakov:2013eya,Bali:2014kia,Tomiya:2019nym}. Later on it is found that the inverse magnetic catalysis is not necessarily associated with the reduction of $T_{pc}$~\cite{DElia:2018xwo,Endrodi:2019zrl}. From lattice QCD studies with larger-than-physical pions one still observes the decreasing of $T_{pc}$ as $eB$ grows, but the inverse magnetic catalysis turns into magnetic catalysis at a sufficiently large pion mass~\cite{DElia:2018xwo,Endrodi:2019zrl}.

In this proceedings we will firstly address the intrinsic connections among the magnetic catalysis, the reduction of $T_{pc}$ and Goldstone pion mass through the Gell-Mann-Oakes-Renner (GMOR) relation as well as from the Ward identity, and then we show a new decreasing behavior observed in the masses of charged pseudo-scalar mesons, and the $qB$ scaling of chiral condensate and neutral pion, finally we will present the first lattice QCD results on the second order fluctuation and correlations of conserved charges at nonzero temperature in strong magnetic fields. The results presented here are based on lattice simulations of $N_f=2+1$ QCD using highly improved staggered fermions with pion mass about 220 MeV and lattice spacing $a\simeq$ 0.117 fm. For studies at zero temperature presented in this proceedings are obtained from $40^3\times96$ and $32^3\times96$ lattices. Details of simulations can be found in \cite{Ding:2020hxw}. While for those at nonzero temperature a fixed scale approach is adopted. This means that we fix $a$ in our study and have temporal extent $N_\tau=16$, 12, 10 and 8 with a spatial extent $N_\sigma=32$ for $T=140$, 168, 210 and 280 MeV, respectively. This corresponds to two temperatures well above $T_{pc}$, one close to $T_{pc}$ and one well below $T_{pc}$ at $eB=0$. Results shown in Section~\ref{sec:vacuum} have been presented in~\cite{Ding:2020hxw}.

\section{Chiral properties of QCD in the vacuum}
In this section we discuss results obtained at $T=0$~\cite{Ding:2020hxw}.
\label{sec:vacuum}
\subsection{Gell-Mann-Oakes-Renner relation and the Ward identity}
The GMOR relation is valid in the chiral limit of quarks at vanishing magnetic field and can be expressed with nonzero pion mass as follows
 ~\cite{GellMann:1968rz}
\begin{equation}
(m_u + m_d) ~\left(\pbp_u + \pbp_d\right)=2 f_\pi^2 M_\pi^2\, (1-\delta_\pi) ,
\label{eq:GMOR}
\end{equation}
where $m_{u,d}$ standing for the up and down quark masses explicitly breaks the chiral symmetry, and $\pbp_{u,d}$ being the up and down quark chiral condensates measures the strength of spontaneous chiral symmetry breaking. $f_\pi$ is the decay constant, $M_\pi$ denotes the Goldstone pion mass and $\delta_\pi$ stands for chiral corrections to the GMOR relation due to a nonzero pion mass. In the presence of nonzero magnetic field the iso-symmetry of up and down quarks is broken due to the different electric charges of up and down quarks. We thus also investigate on the following two relations 
\begin{align}
2m_u~\pbp_u &= f_{\pi^0_u}^2 M_{\pi^0_u}^2 (1-\delta_{\pi^0_u}), \\
2m_d~\pbp_d &= f_{\pi^0_d}^2 M_{\pi^0_d}^2  (1-\delta_{\pi^0_d}),
\label{eq:GMOR}
\end{align}
which hold true at zero magnetic field in the chiral limit with chiral corrections $\delta_{\pi^0_u}=\delta_{\pi^0_d}=0$. Here $f_{\pi^0_{u,d}}$ and $M_{\pi^0_{u,d}}$ are the decay constants and masses corresponding to the up and down flavor components of a neutral pion. At zero magnetic field $f_{\pi^0_{u}}=f_{\pi^0_{d}}=f_{\pi^0}$, $M_{\pi^0_{u}}=M_{\pi^0_{d}}=M_{\pi^0}$ and $\delta_{\pi^0_u}=\delta_{\pi^0_d}=\delta_{\pi^0}$, while at nonzero magnetic field these will not be the case due to the iso-symmetry breaking. 

Given the validity of GOMR relation in the external magnetic field it then suggests that the mechanism for the explicit chiral symmetry breaking by light quark mass is not changed by the magnetic field, and the Goldstone pion mass can be considered as an overall measure of both explicit and spontaneous chiral symmetry breaking. Thus a lighter Goldstone pion mass indicates that the chiral symmetry can be restored at a lower temperature.  From the chiral perturbation theory it has been demonstrated that the GMOR relation holds true in weak magnetic fields and low temperatures~\cite{Gasser:1986vb,Shushpanov:1997sf,Agasian:2001ym}. However, it is not guaranteed that the GMOR relation could hold in strong magnetic fields.

\begin{figure}[htbp]
		\includegraphics[width=0.49\textwidth]{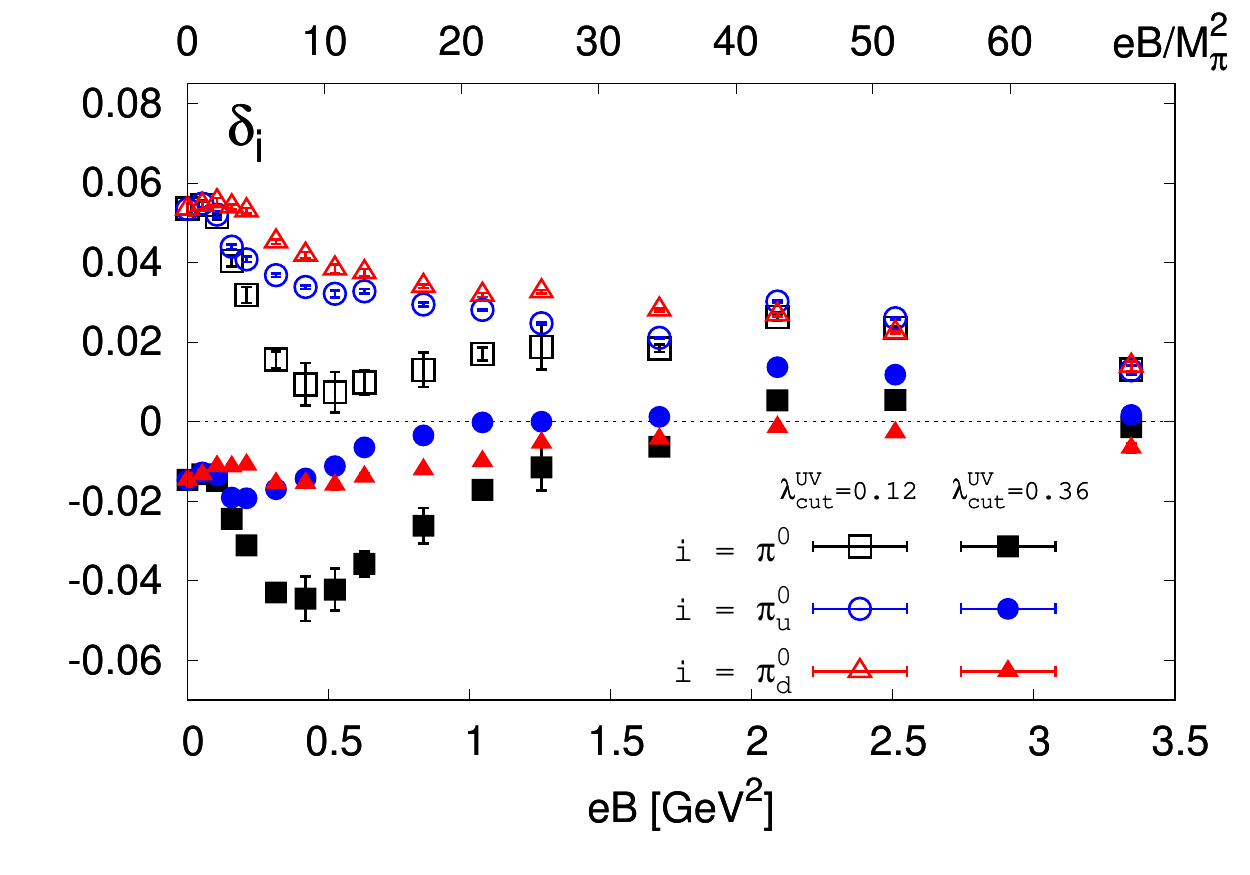}~
	\includegraphics[width=0.49\textwidth]{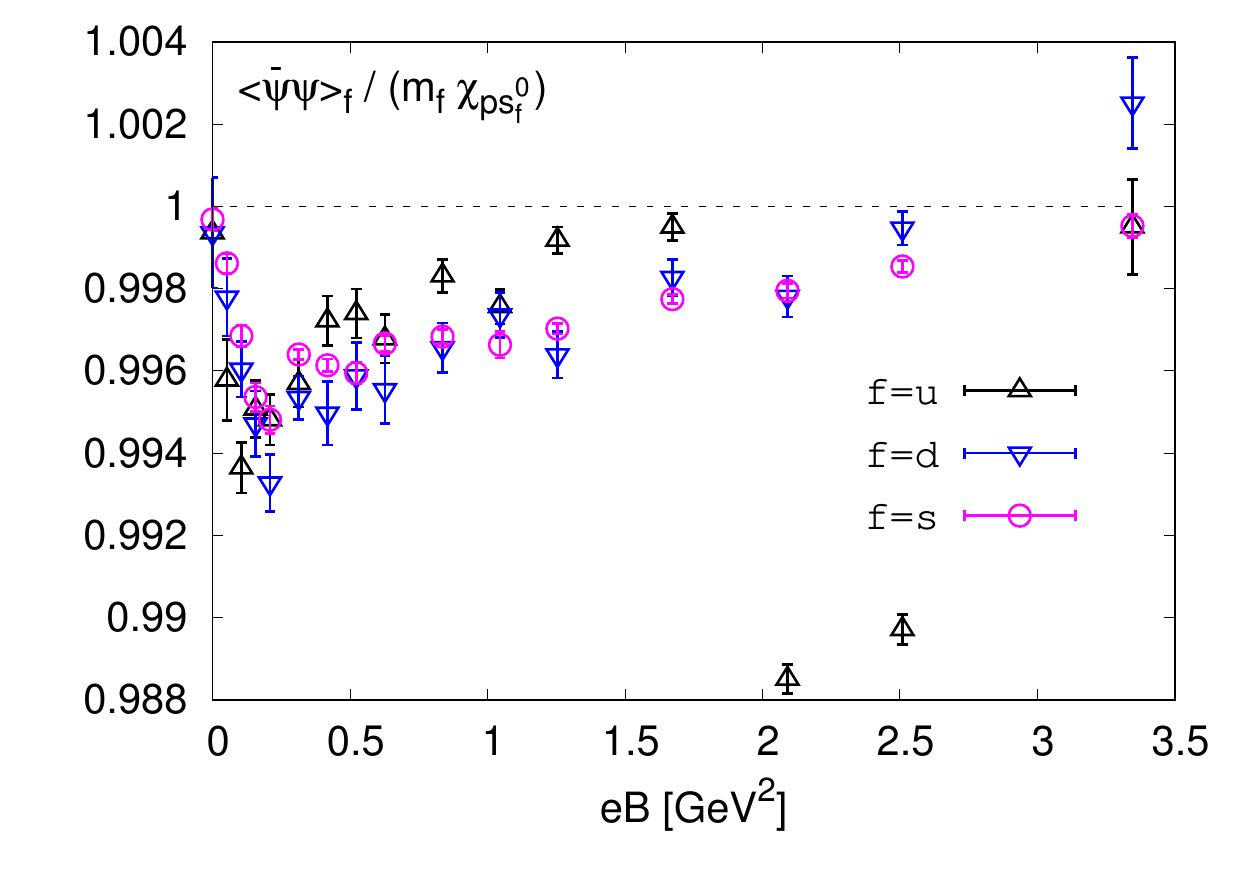}
	\caption{Left: Chiral corrections to the GMOR relation. Right: Ratio of $\pbp_f/(m_f\chi_{\mathrm{ps}^0_f})$ as a function of $eB$ for $f$ as up ($u$), down ($d$) and strange ($s$) quark flavors. The label $M_\pi$ shown in the upper right corner in the left plot and in all the plots through this proceedings stands for pion mass at $eB=0$ in our lattice simulations, $i.e.$ $M_\pi(eB=0)=220$ MeV. Figures are taken from~\cite{Ding:2020hxw}.}
	\label{fig:GMORandWI}
\end{figure}

In the left panel of Fig.~\ref{fig:GMORandWI} we show the chiral corrections to the GMOR relation $|\delta|$ is at most 6\% at $eB=0$ as obtained from different subtractions of the UV part of the light quark chiral condensate. As $eB$ grows $|\delta|$ gets closer to zero. This suggests that the GMOR relation approximately holds in the currently explored $eB$ region.

In the right panel of Fig.~\ref{fig:GMORandWI} we show whether the Ward identity of $\pbp_{u,d}=m_{u,d}\chi_{\pi^0_{u,d}}$ holds in the nonzero magnetic field. It can be seen that the ratio  $\pbp_{u,d}/(m_{u,d}\chi_{\pi^0_{u,d}})$ deviates from unity in the nonzero magnetic field by at most 1.2\%. Since $\chi_{\pi^0_{u,d}}$ is the space-time sum of the meson correlation functions of $\pi^0_{u,d}$ it is dominated by the exponential decay of $M_{\pi^0_{u,d}}$ at large distances. Thus a smaller $M_{\pi^0_{u,d}}$ ($cf.$ Fig.~\ref{fig:PSMass}) is consistent with larger values of both $\chi_{\pi^0_{u,d}}$ and $\pbp_{u,d}$ as $eB$ grows, and it is also in accord with the reduction of $T_{pc}$ in the nonzero magnetic field.

\subsection{Pseudo-scalar meson masses and $qB$ scaling}
\begin{figure}[!htbp]
	\includegraphics[width=0.49\textwidth]{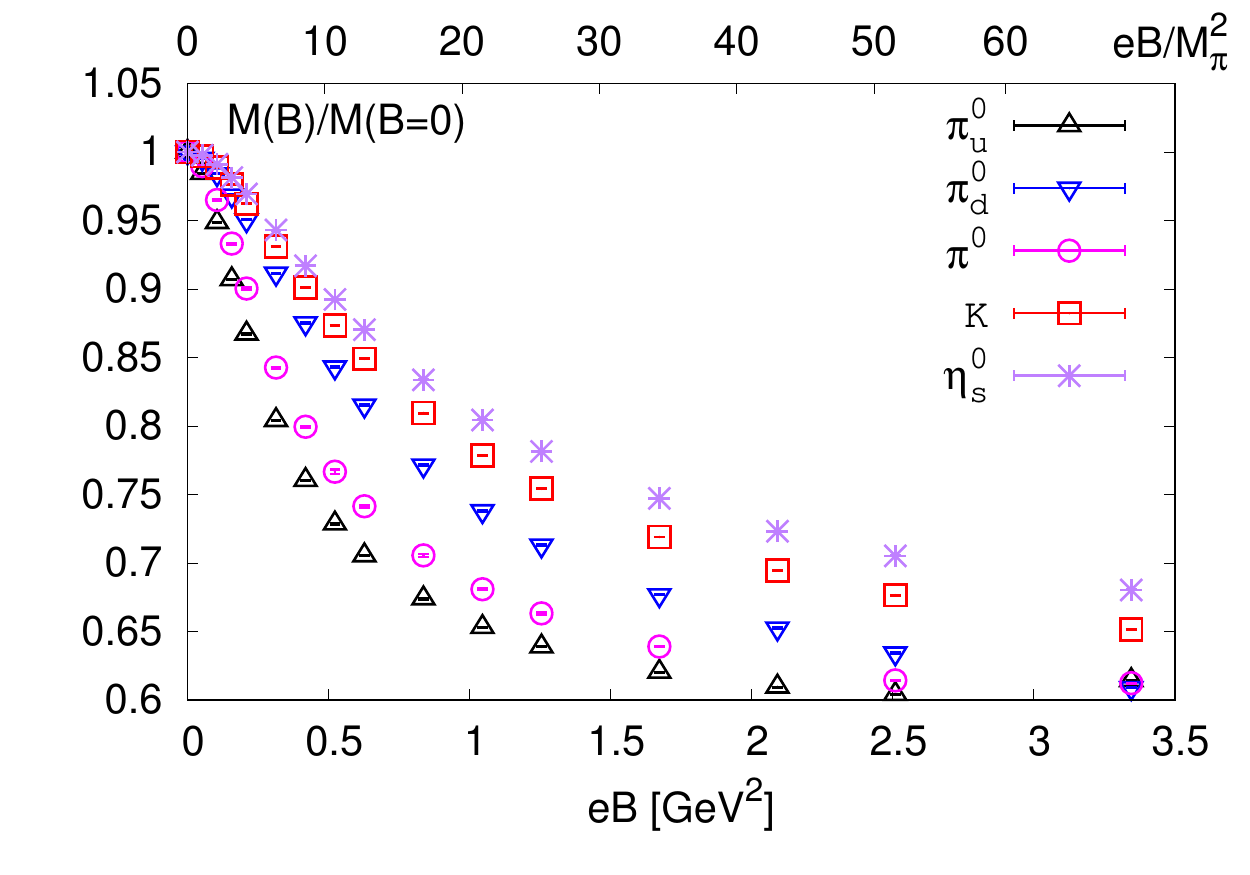}
		\includegraphics[width=0.49\textwidth]{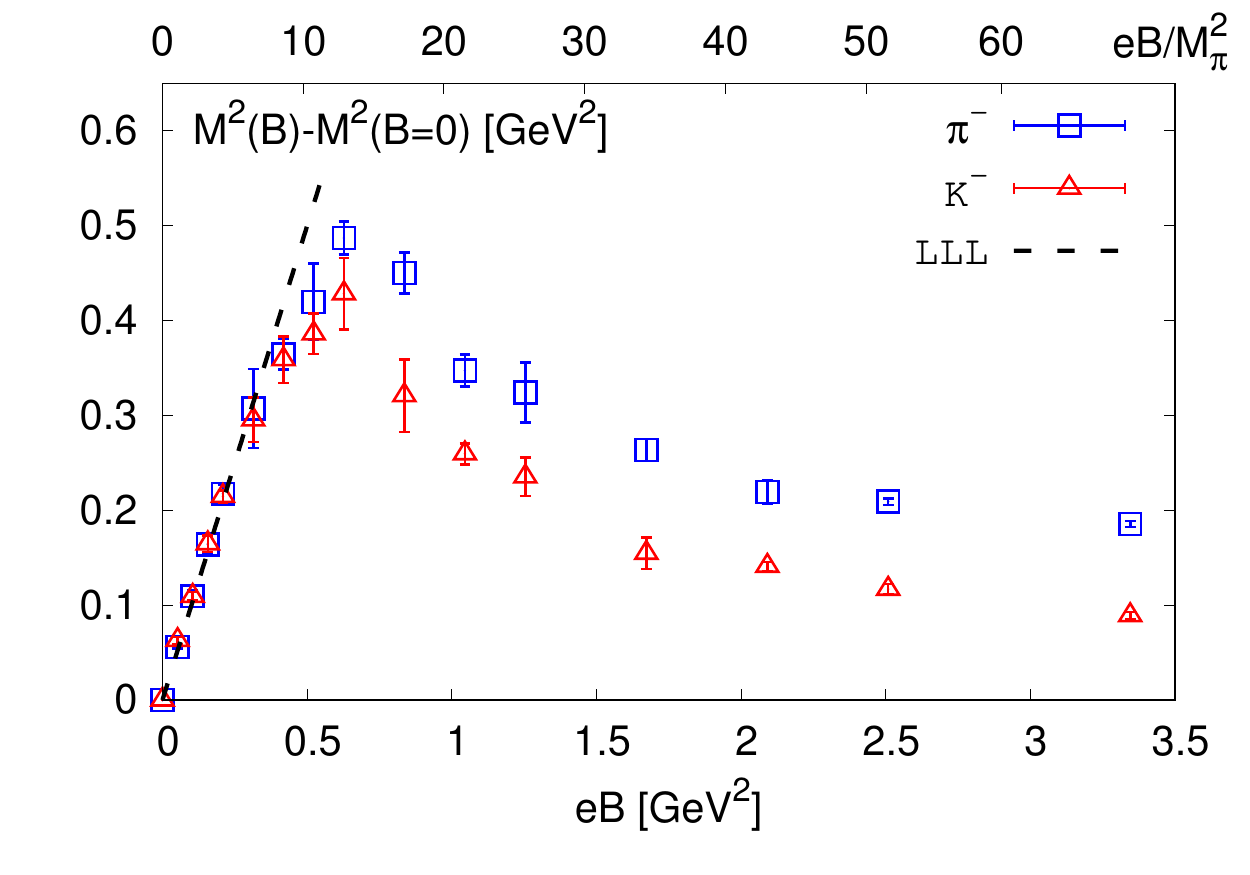}
	\caption{Left: Masses of $\pi^0_{u,d}$, $\pi^0$, $K$, $\eta^0_s$ normalized by their corresponding values at $eB$=0 as a function of $eB$. Right: $M^2(eB)-M^2(eB=0)$ for charged pseudo-scalar mesons. Figures are taken from Ref.~\cite{Ding:2020hxw}.}
	\label{fig:PSMass}
\end{figure}

In left panel of Fig.~\ref{fig:PSMass} we show the $eB$ dependence of neutral pseudo-scalar meson masses normalized by their corresponding values at vanishing magnetic field. We see that masses of $\pi^0_{u,d}$, $\pi^0$, $K^0$ and $\eta_s^0$ monotonously decrease with $eB$ and seem to saturate at $eB\gtrsim2.5$ GeV$^2$. A lighter Goldstone pion, i.e. neutral pion in stronger magnetic field thus suggests a lower transition temperature, and this is in analogy to the case with lighter quark mass at zero magnetic field~\cite{Ding:2019prx,Kaczmarek:2020err,Ding:2020zpp}.

In right panel of Fig.~\ref{fig:PSMass} we show the differences of $\pi^-$ and $K^-$ mass squared from their corresponding values at vanishing magnetic field as a function of $eB$. At $eB\lesssim0.3$ GeV$^2$ the lattice data can be well described by the lowest Landau level (LLL) approximation as denoted by the dashed line. This suggests that both $\pi^-$ and $K^-$ behave as point-like particles. At larger values of $eB$ the lattice data starts to deviate from the LLL approximation and then decreases with $eB$ at a turning point of $eB\approx0.6$ GeV$^2$. We have checked the volume dependence at $eB\simeq1.67$ GeV$^2$ using a larger lattice of $40^3\times96$, and find that the volume dependence is negligible. Thus the decreasing behavior of $M_{\pi^-}$ and $M_{K^-}$ at $eB\gtrsim0.6$ GeV$^2$ should be robust. This indicates that the internal structure of $\pi^-$ and $K^-$ are probed by the magnetic field at $eB\gtrsim0.3$ GeV$^2$. This decreasing behavior of charged pseudo-scalar mesons as $eB$ grows is new, and it could be due to the effects arising from dynamics quarks and large magnetic field we simulated in our study compared to previous studies in~\cite{Bali:2011qj,Luschevskaya:2014lga,Bali:2017ian}.

\begin{figure}[tbph]
	\centering
	\includegraphics[width=0.49\textwidth]{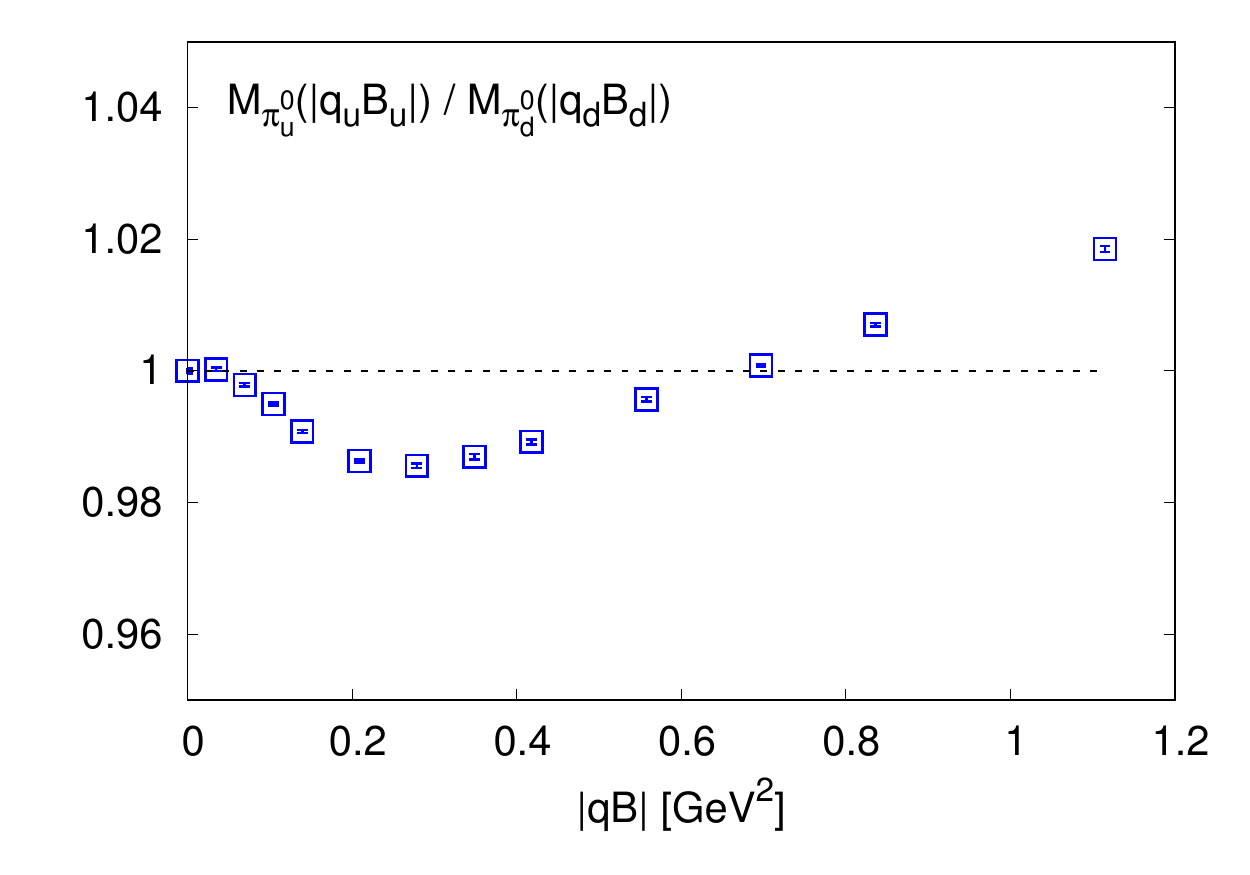}
    \includegraphics[width=0.49\textwidth]{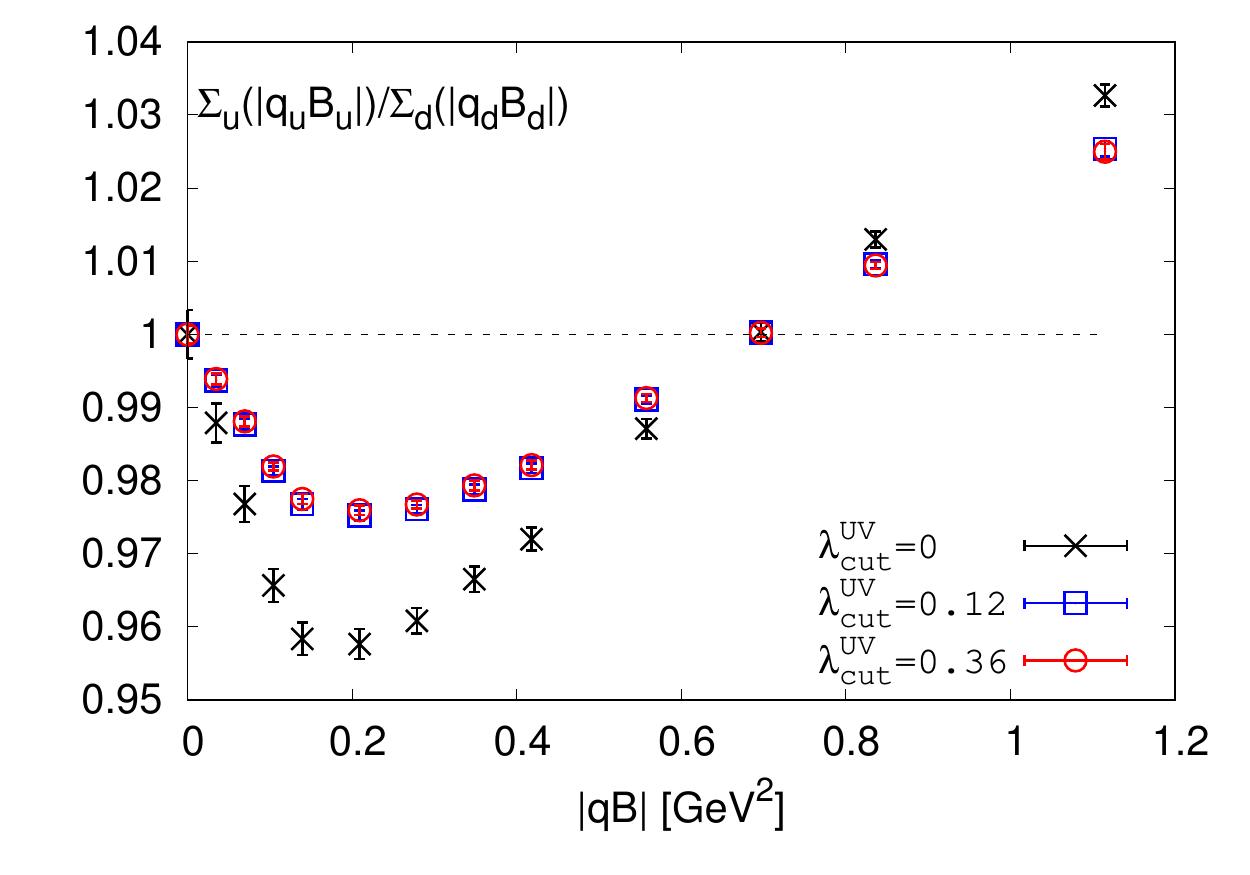}
	\caption{ $qB$ scaling seen in the up and down quark components of neutral pion mass (left) and up and down quark chiral condensates (right). Figures are taken from Ref.~\cite{Ding:2020hxw}.}
	\label{fig:qBscaling}
\end{figure}

In the left panel of Fig.~\ref{fig:qBscaling}
 we show the ratio $M_{\pi^0_u}(|q_uB_u|)/M_{\pi^0_d}(|q_dB_d|)$ as a function of $|qB|$. Here $q_u$ and $q_d$ stand for the electric charges of $u$ and $d$ quarks, and $B_{u}$ and $B_{d}$ are different values of $B$ which makes $|qB|\equiv|q_uB_u|=|q_dB_d|$. The ratio is very close to 1. We call this $qB$ scaling, according to which the ratio should be exactly 1 in the quenched limit. The $qB$ scaling can also be found in the up and down quark chiral condensate as shown in the right plot of Fig.~\ref{fig:qBscaling}. These findings suggest that the effects from dynamic quarks are negligible in our case with $M_\pi(eB=0)=220$ MeV.

\section{Fluctuations of and correlations among conserved charges}
Fluctuations of and correlations among conserved charges are useful probes to study the QCD phase structure.
They have been extensively studied in lattice QCD at vanishing magnetic field~\cite{Ding:2015ona,Ding:2020rtq,Guenther:2020vqg}. We extend the study of these quantities to the case in strong magnetic fields which could also be of interests~\cite{Fu:2013ica,Bhattacharyya:2015pra,Fukushima:2016vix,Mohapatra:2017zrj,Ferreira:2018pux}. Following the same procedure as presented in e.g.~\cite{Bazavov:2020bjn},  we compute the fluctuations of the conserved charges and their correlations by taking the derivatives of pressure $p$ with respect to the chemical potentials from lattice calculation directly,
\begin{align}
\frac{\chi _ { i j k } ^ { u d s }}{T^2}  &= \frac { \partial ^ { i + j + k } p / T ^ { 4 } } { \partial \left( \mu _ { u } / T \right) ^ { i } \partial \left( \mu _ { d } / T \right) ^ { j } \partial \left( \mu _ { s } / T \right) ^ { k } }\Big|_{\mu_u=\mu_d=\mu_s=0},\\
\frac{\chi _ { i j k } ^ { B Q S }}{T^2}  &= \frac { \partial ^ { i + j + k } p / T ^ { 4 } } { \partial \left( \mu _ { B } / T \right) ^ { i } \partial \left( \mu _ { Q } / T \right) ^ { j } \partial \left( \mu _ { S } / T \right) ^ { k } }{\Big|}_{\mu_B=\mu_Q=\mu_S=0}.
\end{align}
Here $\mu_{u,d,s}$ stand for up, down and strange quark chemical potentials while $\mu_{B,Q,S}$ denote baryon number, electric charge and strangeness chemical potentials. In our study we consider the case of $i+j+k=2$, $i.e.$ second order fluctuations and correlations.

\begin{figure}[tbph]
	\centering
	\includegraphics[width=0.49\textwidth]{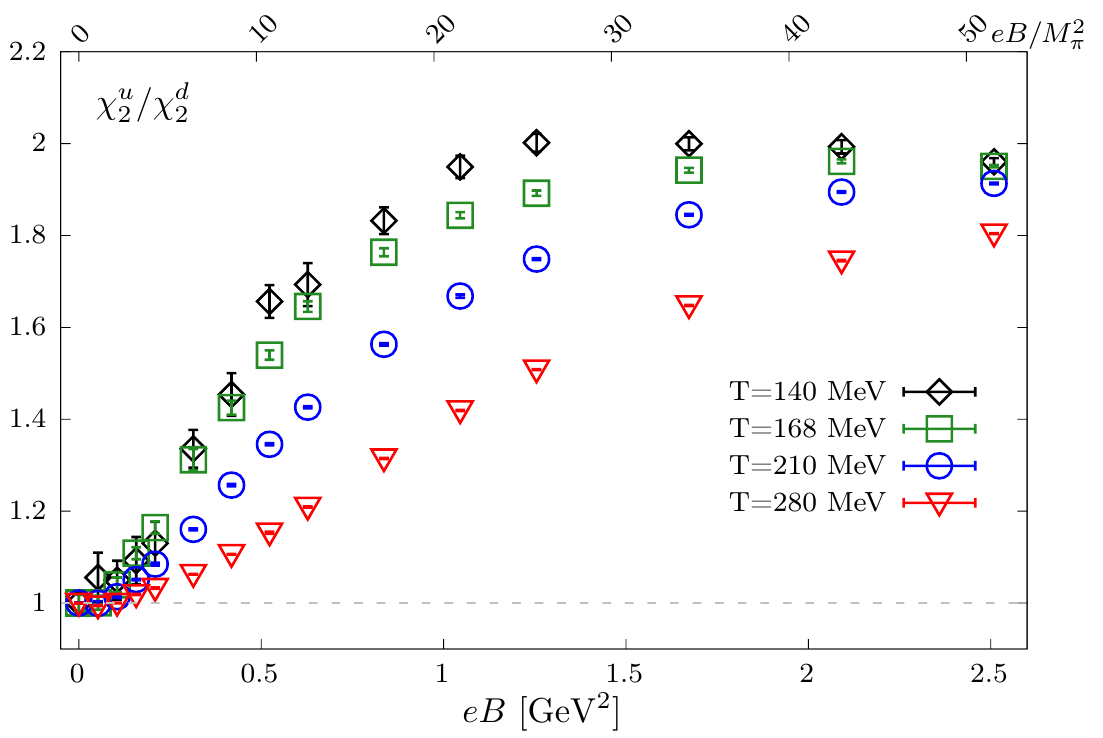}
 	\includegraphics[width=0.49\textwidth]{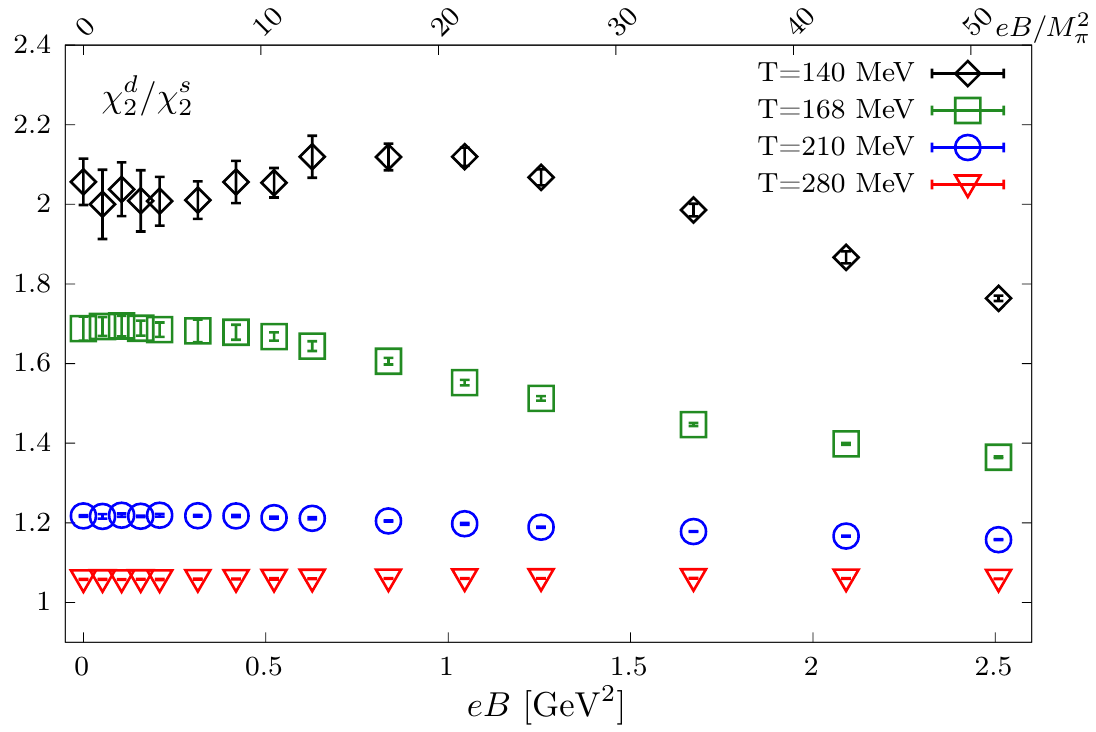}
	\caption{Left: Ratio of up quark number susceptibility $\chi_2^u$ to down quark number susceptibility $\chi_2^d$. Right: Ratio of $\chi_2^d$ to strange quark number susceptibility $\chi_2^s$.}
	\label{fig:chi2uds}
\end{figure}

In the left panel of Fig.~\ref{fig:chi2uds} we show $\chi_2^u/\chi_2^d$ as a function of $eB$ at temperatures below, close and above the transition temperature at $eB=0$. This ratio $\chi_2^u/\chi_2^d$ is 1 at $eB=0$ and its deviation from 1 at nonzero values of $eB$ is due to the iso-symmetry breaking of up and down quarks. And the iso-symmetry breaking effects are observed to be larger at lower temperatures. We also want to check how the effects induced by $eB$ are manifested in the ratio $\chi_2^d/\chi_2^s$ with down quark and strange quark having same electric charges. This is shown in the right panel of Fig.~\ref{fig:chi2uds}. It can be seen that at each temperature the ratio generally becomes smaller as $eB$ grows, and the magnitude of change is much smaller compared to the case for $\chi_2^u/\chi_2^d$.

\begin{figure}[tbph]
	\centering
		\includegraphics[width=0.49\textwidth]{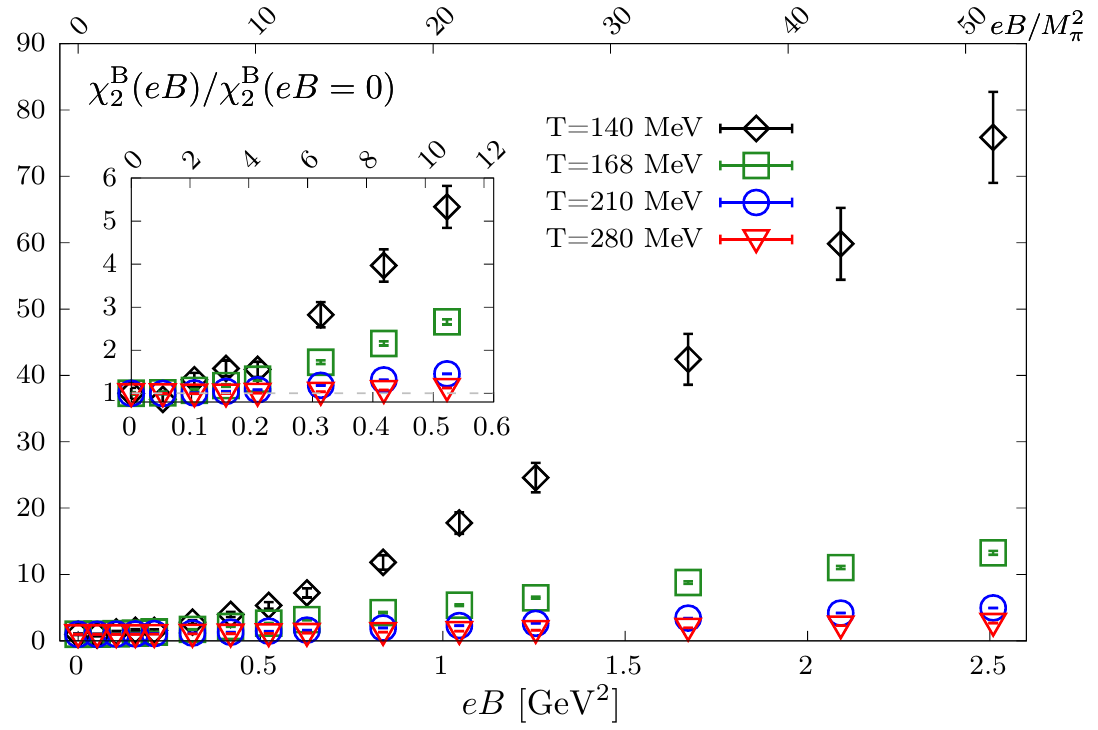}
				\includegraphics[width=0.49\textwidth]{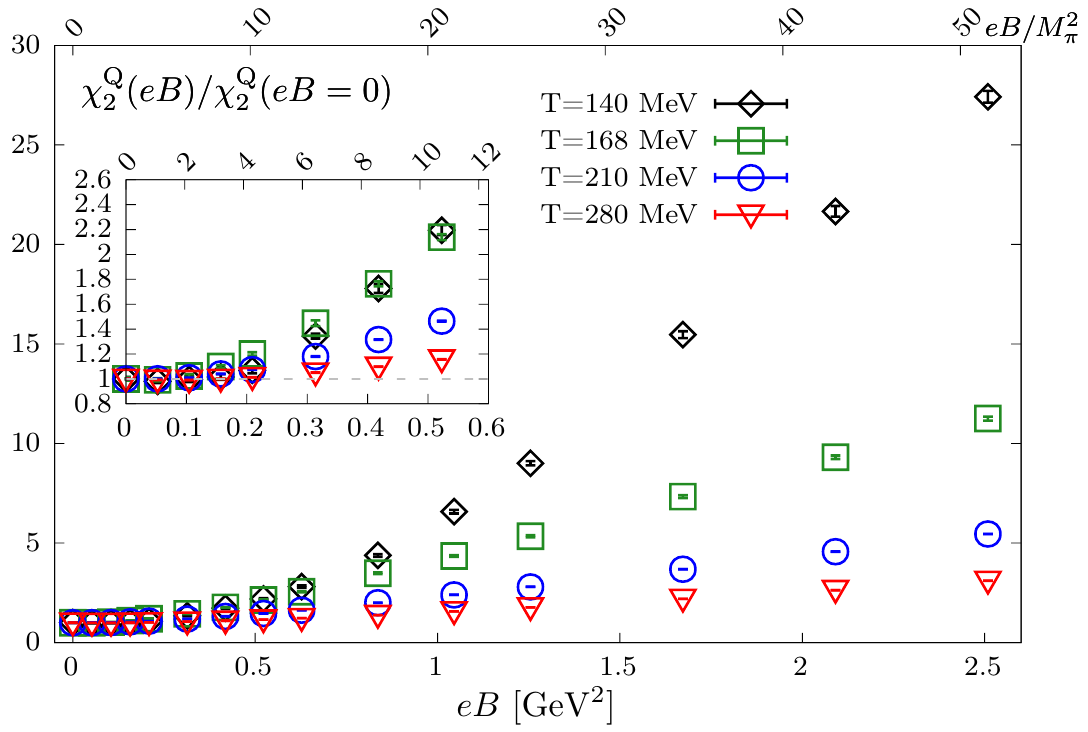}
	\caption{Left: $\chi_2^{\mathrm{B}}$ divided by its value at $eB=0$ as a function of $eB$. The inset shows a blow-up plot for $eB\in[0, 0.6]$ GeV$^2$. Right: Same as the left panel but for $\chi_2^{\mathrm{Q}}$.}
	\label{fig:chi2BQ}
\end{figure}
Now we look into the fluctuations of conserved charges in the presence of magnetic field.
We show $\chi_2^{\mathrm{B}}(eB)/\chi_2^{\mathrm{B}}(eB=0)$ and  $\chi_2^{\mathrm{Q}}(eB)/\chi_2^{\mathrm{Q}}(eB=0)$ as functions of $eB$ in the left and right panels of Fig.~\ref{fig:chi2BQ}, respectively. It can be observed that both $\chi_2^{\mathrm{B}}$ and $\chi_2^{\mathrm{Q}}$ monotonously increase as $eB$ grows at all four different temperatures. The magnitude of increasing becomes larger at lower temperatures. At $T$=140 MeV and in the strongest magnetic field we simulated $\chi_2^B$ and $\chi_2^Q$ becomes about 75 and 28 times of their corresponding values at $eB=0$, respectively. At the highest temperature we have, the changes induced by the magnetic field is much smaller, $i.e.$ about 3 times for both $\chi_2^{\mathrm B}$ and $\chi_2^{\mathrm Q}$. We also show blow-up plots for $eB\in[0,0.6]$ GeV$^2$ as insets in both panels. This $eB$ region corresponding to 0-12 $M_\pi^2(eB=0)$ could be the strength of magnetic field reached at the initial stage of heavy ion collision experiments~\cite{Skokov:2009qp}. At T=140 MeV we see that $\chi_2^{\mathrm{B}}$ ($\chi_2^{\mathrm{Q}}$) at $eB\approx11$ $M_\pi^2(eB=0)$ is about 6 (2) times its value at $eB=0$.

In the high temperature (free) limit,  most of these quantities $\chi _ { i j k } ^ { B Q S }$ divided by $eB$ scales with $\sqrt{eB}/T$. For instance the expression of $\chi_2^{\mathrm B}$ is written as follows 
\begin{align}
\frac{9\pi^2}{4}\frac{{\chi}_2^\mathrm{B}}{eB} =  \frac{1}{2} + \frac{\sqrt{2eB/3}}{T} \sum_{l=1}^{\infty} \sqrt{l} \,\sum_{k=1}^\infty (-1)^{k+1}k \times  
&\Bigg[\sqrt{2}\cdot{\rm K}_1\left(\frac{2k\sqrt{eBl/3}}{T}\right) \nonumber\\
& + {\rm K}_1\left(\frac{k\sqrt{2eBl/3}}{T}\right)\Bigg] ,
\end{align}
where $l$ denotes the Landau levels, $k$ is the sum index in Taylor expansion series and $K_1$ is the modified Bessel function of the second kind. For some certain ratios, e.g. $-3\chi_{11}^{\mathrm BS}/\chi_2^{\mathrm{S}}$ is always 1 in the high temperature limit at both vanishing and nonzero magnetic field. 

\begin{figure}[tbph]
	\centering
	\includegraphics[width=0.49\textwidth]{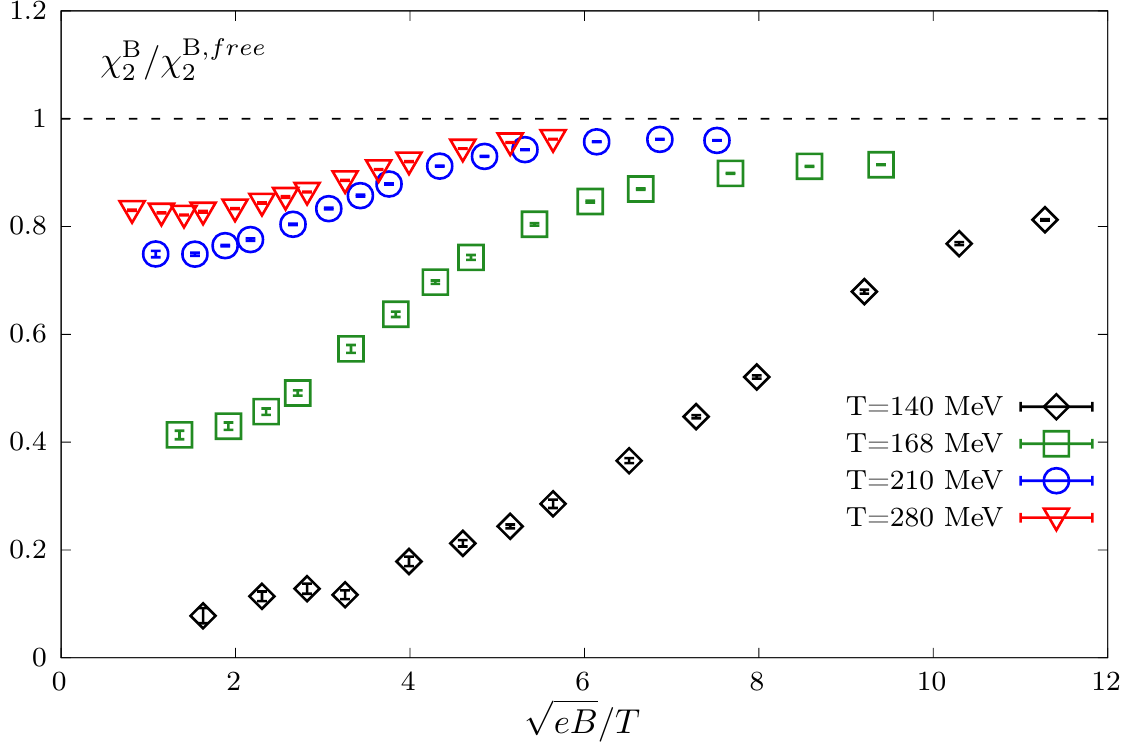}
	\includegraphics[width=0.49\textwidth]{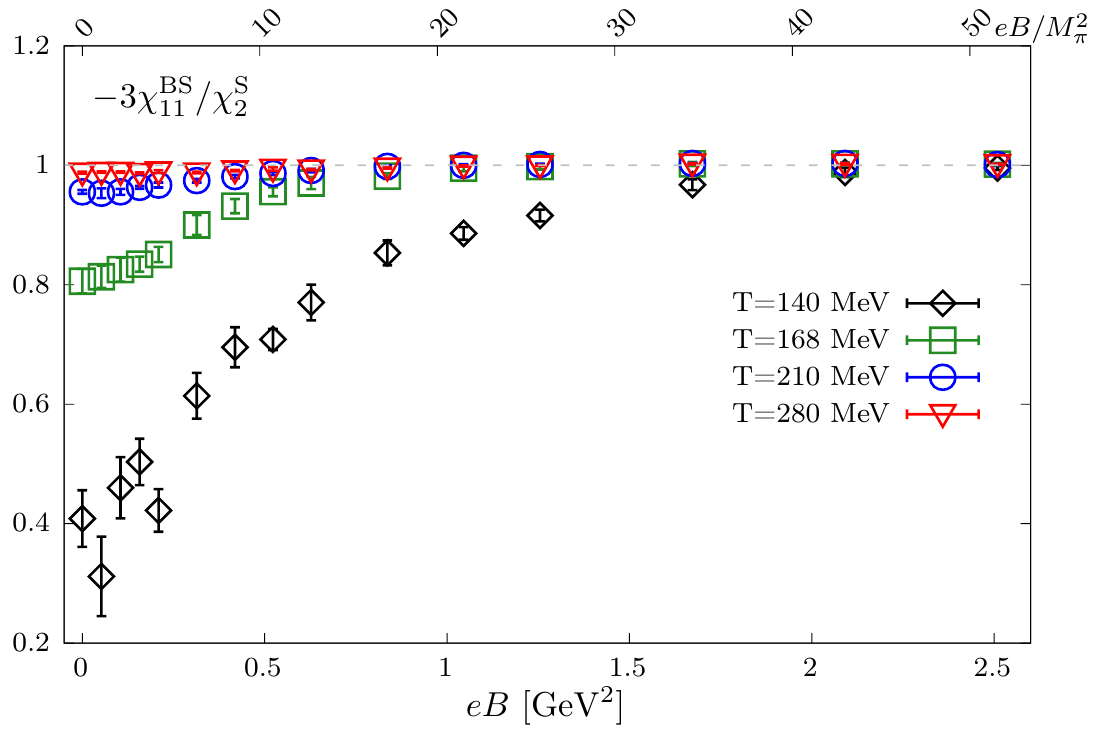}
	\caption{Left: $\chi_2^\mathrm{B}$ divided by its value in the free limit $\chi_2^{\mathrm{B},free}$  as a function of $\sqrt{eB}/T$.  Right: $-3\chi_{11}^{BS}/\chi_2^S$ as a function of $eB$. }
	\label{fig:chi2free}
\end{figure}

In the left panel of Fig.~\ref{fig:chi2free} we show $\chi_2^{\mathrm{B}}$ divided by its value in the free limit as a function of $\sqrt{eB}/T$ at four different temperatures. It can be clearly seen that $\chi_2^{\mathrm{B}}$ becomes closer to its value in the free limit as $eB$ grows at all four temperatures. And it can also been observed that a stronger magnetic field is needed to bring $\chi_2^B$ to its value in the free limit at a lower temperature.  Similar conclusion can be drawn from $eB$ and temperature dependences of $-3\chi_{11}^{BS}/\chi_2^S$, whose value in the free limit is always 1, in the right panel of Fig.~\ref{fig:chi2free}. The observation from Fig.~\ref{fig:chi2free} can be understood as the magnetic field catalyzes the phase transition.
\begin{figure}[tbph]
	\centering
	\includegraphics[width=0.47\textwidth]{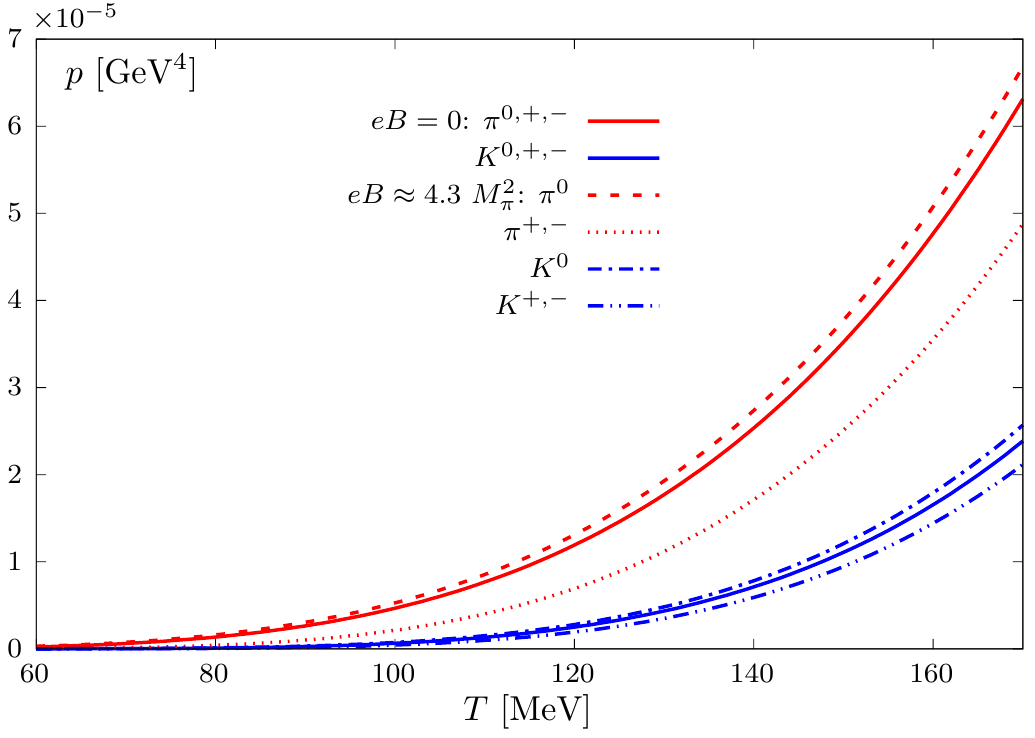}
	\includegraphics[width=0.47\textwidth]{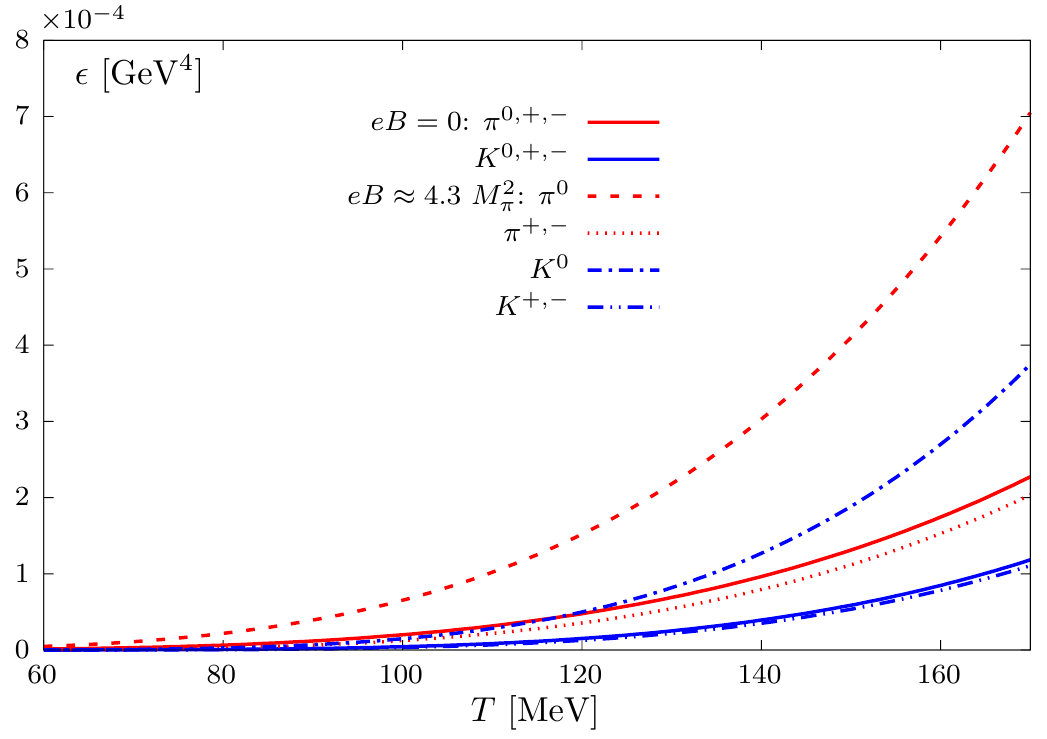}
	\caption{Pressure (left) and energy density (right) contributed from individual pseudo-scalar mesons obtained from HRG at $eB=0$ and $eB=0.21$ GeV$^2\approx 4.3$ $M_\pi^2(eB=0)$. Solid lines denote results obtained at $eB=0$ while the other lines represent results obtained at $eB\approx 4.3$ $M_\pi^2(eB=0)$ in both panels. The meson masses used in the HRG model are obtained from Fig.~\ref{fig:PSMass}.}
	\label{fig:HRGpe}
\end{figure}

At low temperatures the hadron resonance gas (HRG) model is supposed to give good description of the fluctuations of conserved charges~\cite{Bazavov:2012jq,Bazavov:2014xya,1828476,Ding:2020rtq}. In presence of magnetic field the situation becomes more complex as the hadron spectrum gets modified by the magnetic field~\cite{Endrodi:2013cs,Endrodi:2019whh,Ding:2020jui,Ding:2020hxw}. Thus to reproduce the quantities shown in Fig.~\ref{fig:chi2BQ} and Fig.~\ref{fig:chi2free} correct hadron mass spectrum in the magnetic field is needed. Here we rather show the individual contributions from several pseudo-scalar mesons to the pressure and energy density in the left and right panels of Fig.~\ref{fig:HRGpe} at $eB=0$ and $0.21$ GeV$^2$. It can be clearly seen that the contribution from $\pi^0$ always dominates in particular to the energy density.

\section{Conclusions}
The results presented in this proceedings are based on the lattice simulations of $N_f=2+1$ QCD with $M_\pi(eB=0)=220$ MeV using Highly Improved Staggered fermions. At zero temperature we find that the Gell-Mann-Oakes-Renner relation and Ward identity hold true for neutral pions in strong magnetic fields. The reduction of neutral pion mass thus is consistent with the magnetic catalysis, and the reduction of $T_{pc}$ in the magnetic field. The latter consistency is also indicated from the dominance of $\pi^0$'s contribution to the energy density in the magnetic field obtained form the HRG model. We also find that the $qB$ scaling holds true for up and down quark components of neutral pion mass as well as light quark chiral condensates. This suggests that the effects from dynamic quarks in QCD with $M_\pi=220$ MeV on the $qB$ scaling are negligible. For the charged pseudo-scalar meson mass a novel decreasing behavior is observed at $eB\gtrsim 0.6$ GeV$^2$. At nonzero temperature the increasing of $\chi_2^\mathrm{B}$ and $\chi_2^\mathrm{Q}$ compared to their corresponding values at $eB=0$ is significant in the presence of magnetic field. This could be observed in heavy ion collision experiments if the strong magnetic field produced in the peripheral collisions lives sufficiently long. The observation that both $\chi_2^\mathrm{B}$ and $-3\chi_{11}^{\mathrm{BS}}/\chi_2^{\mathrm{S}}$ get closer to their free limit at lower temperatures in the presence of magnetic field is in accord with the fact that $T_{pc}$ decreases with $eB$.

\section*{Acknowledgements}
We thank Toru Kojo for interesting discussions. 
This material is based upon work supported by the National Natural Science
Foundation of China under Grants Nos. 11535012, 11775096 and 11947237, and the RIKEN Special Postdoctoral Researcher program and JSPS KAKENHI Grant
Number JP20K14479. Computations for this work were carried out on the GPU clusters of the Nuclear
Science Computing Center at Central China Normal University (NSC$^3$), Wuhan, China.

\bibliographystyle{JHEP}
\bibliography{refs}

\end{document}